\documentclass[
preprint,
superscriptaddress,
nofootinbib,
amsmath,
amssymb,
aps,
pra,
floatfix,
]{revtex4-2}

\usepackage{graphicx}      % Include figure files
\usepackage{dcolumn}      % Align table columns on decimal point
\usepackage{bm}         % bold math
\usepackage{siunitx}
\DeclareSIUnit{\ppm}{ppm}
\DeclareSIUnit{\pixel}{px}
\DeclareSIUnit{\bar}{bar}
\usepackage{fp}
\newcommand{\AlGaAs}[1][0.929]{\FPeval{\result}{round(1-#1,3)}Al\textsubscript{#1}Ga\textsubscript{\result}As}
\newcommand{\AlGaAsx}{Al\textsubscript{x}Ga\textsubscript{1-x}As}
\usepackage[inline]{enumitem}
\begin{document}

%\preprint{APS/123-QED}

\title{Simultaneous Measurement of Mid-Infrared Refractive Indices in Thin-Film Heterostructures: Methodology and Results for GaAs/AlGaAs}

\author{Lukas W. Perner}
  \email{lukas.perner@univie.ac.at}
  \affiliation{Christian Doppler Laboratory for Mid-IR Spectroscopy and Semiconductor Optics, Faculty Center for Nano Structure Research, Faculty of Physics, University of Vienna, Boltzmanngasse 5, 1090 Vienna, Austria.}
  \affiliation{Vienna Doctoral School in Physics, University of Vienna, Boltzmanngasse 5, 1090 Vienna, Austria.}
\author{Gar-Wing Truong}
  \affiliation{Thorlabs Crystalline Solutions, 114 E Haley St., Suite G, Santa Barbara, California 93101, USA}
\author{David Follman}
  \affiliation{Thorlabs Crystalline Solutions, 114 E Haley St., Suite G, Santa Barbara, California 93101, USA}
\author{Maximilian Prinz}
  \affiliation{Christian Doppler Laboratory for Mid-IR Spectroscopy and Semiconductor Optics, Faculty Center for Nano Structure Research, Faculty of Physics, University of Vienna, Boltzmanngasse 5, 1090 Vienna, Austria.}
\author{Georg Winkler}
  \affiliation{Christian Doppler Laboratory for Mid-IR Spectroscopy and Semiconductor Optics, Faculty Center for Nano Structure Research, Faculty of Physics, University of Vienna, Boltzmanngasse 5, 1090 Vienna, Austria.}
\author{Stephan Puchegger}
  \affiliation{Faculty Center for Nano Structure Research, Faculty of Physics, University of Vienna, Boltzmanngasse 5, 1090 Vienna, Austria.}
\author{Garrett D. Cole}
  \affiliation{Thorlabs Crystalline Solutions, 114 E Haley St., Suite G, Santa Barbara, California 93101, USA}
\author{Oliver H. Heckl}
  \affiliation{Christian Doppler Laboratory for Mid-IR Spectroscopy and Semiconductor Optics, Faculty Center for Nano Structure Research, Faculty of Physics, University of Vienna, Boltzmanngasse 5, 1090 Vienna, Austria.}

\date{\today}

\begin{abstract}
We present our results for simultaneous measurement of the refractive indices of gallium arsenide (GaAs) and aluminum gallium arsenide (\AlGaAsx{}) in the spectral region from \SIrange{2.0}{7.1}{\micro\meter} (\SIrange{5000}{1400}{\per\cm}). We obtain these values from a monocrystalline superlattice Bragg mirror of excellent purity (background doping $\leq \SI{1e-14}{\per\cm\cubed}$), grown via molecular beam epitaxy. To recover the refractive indices over such a broad wavelength range, we fit a dispersion model for each material.
In a novel combination of well-established methods, we measure both a photometrically accurate transmittance spectrum of the Bragg mirror via Fourier-transform infrared spectrometry and the individual physical layer thicknesses of the structure via scanning electron microscopy. To infer the uncertainty of the refractive index values, we estimate relevant measurement uncertainties and propagate them via a Monte-Carlo method. This highly-adaptable approach conclusively yields propagated relative uncertainties on the order of $10^{-4}$ over the measured spectral range for both GaAs and \AlGaAs{}. The fitted model can also approximate the refractive index for MBE-grown \AlGaAsx{} for $0 \leq x \leq 1$.
Both these updated values and the measurement approach will be essential in the design, fabrication, and characterization of next-generation active and passive optical devices in a spectral region that is of high interest in many fields, e.g., laser design and cavity-enhanced spectroscopy in the mid-infrared spectral region.
\end{abstract}

\maketitle

\section{Introduction}
Heterostructures based on gallium arsenide (GaAs) and aluminum gallium arsenide (\AlGaAsx{}, where $x$ denotes the AlAs mole fraction) are paramount in the design and production of a multitude of active and passive (electro-)optical devices ranging from light sources, such as vertical-cavity surface-emitting lasers (VCSELs)~\cite{2001Vertical-CavityLasers}, superluminescent diodes~\cite{Lin1997ExtremelyDiodes} and quantum cascade lasers (QCLs)~\cite{Sirtori1998GaAs/AlxGa1-xAsLasers}, plus detection devices, such as quantum cascade detectors (QCDs), quantum-well infrared photodetectors (QWIPs)~\cite{Hofstetter2010Mid-infraredPyrometry}, and megapixel infrared camera sensors based on the QWIP technology~\cite{Gunapala2005DevelopmentCameras}, to highly-reflective (HR) distributed Bragg reflectors (DBRs)~\cite{vanderZiel1975MultilayerEpitaxy,vanderZiel1976InterferenceAlAs-GaAs,Heiss2001EpitaxialApplications}, semiconductor saturable absorber mirrors (SESAMs)~\cite{Keller1996SemiconductorLasers,Keller2010UltrafastSight}, and HR optomechanical resonators~\cite{Cole2008MonocrystallineRegime}.
Owing to mature fabrication technologies, notably molecular beam epitaxy (MBE)~\cite{Pohl2020EpitaxySemiconductors}, GaAs/\AlGaAsx{}-based devices find extensive applications in the near-infrared (NIR) and mid-infrared (MIR) wavelength range. For example, a recently-developed technology allows the transfer of MBE-grown GaAs/\AlGaAsx{} HR DBRs to curved optical substrates, making the material system relevant for applications from gravitational wave detection~\cite{Barr2012LIGOT1200046-v1,cole_substrate-transferred_2023} to ultra-narrow linewidth laser stabilization in the NIR~\cite{Cole2013TenfoldCoatings}. This substrate-transfer method was extended to the MIR wavelength range~\cite{Cole2016, Winkler2021Mid-infraredPpm, Truong2022Transmission-dominated000}, enabling ultra-low excess loss at MIR wavelengths up to \SI{4.5}{\um}, making substrate-transferred GaAs/\AlGaAsx{}-based DBRs a key technology for advances in cavity-enhanced MIR spectroscopy applications~\cite{Gagliardi2014Cavity-EnhancedSensing,Bjork2016DirectKinetics}.

Despite the extensive application of GaAs/\AlGaAsx{} in optical devices, there is a lack of recent accurate and precise data for its optical properties, especially refractive index $n$, in the MIR wavelength range (\SI{>2}{\um}). The existing literature values for these materials are largely based on the characterization of bulk samples rather than MBE-grown multilayers. Probing such samples, along with often incomplete reporting of measurement conditions, leads to substantial differences in refractive index, especially when comparing samples grown by different methods~\cite{Sell1974ConcentrationEV}. This discrepancy, likely due to variations in growth, background doping, and purity levels (see Fig.\ref{fig:final_n_results}), calls for updated literature values for GaAs, \AlGaAsx{}, among other materials~\cite{Kaplan1998FourierRefractometry}.

For \AlGaAsx{}, which is used with a wide variety of mole fractions $x$, the latest semi-empirical model that allows for an arbitrary $x$ was published in 1974~\cite{Afromowitz1974}, where model parameters were obtained from a fit to measurement data obtained in the NIR range (approx. from \SIrange{680}{1030}{\nm}) on samples grown via liquid-phase epitaxy (LPE)~\cite{Casey1974RefractiveEV}. More recent studies for GaAs and \AlGaAsx{} exist~\cite{Skauli2003ImprovedOptics,Palmer2002Mid-infraredAlAs,Papatryfonos2021RefractiveRegion}. However, these either focus on the NIR spectral region or are of insufficient accuracy for many of the above applications, which rely on MBE-grown GaAs/\AlGaAsx{} due to discrepancies in measurement conditions and sample material (e.g., when using $n$ measured for bulk material in the design of thin film heterostructures).

\begin{figure*}
    \includegraphics[width=\columnwidth]{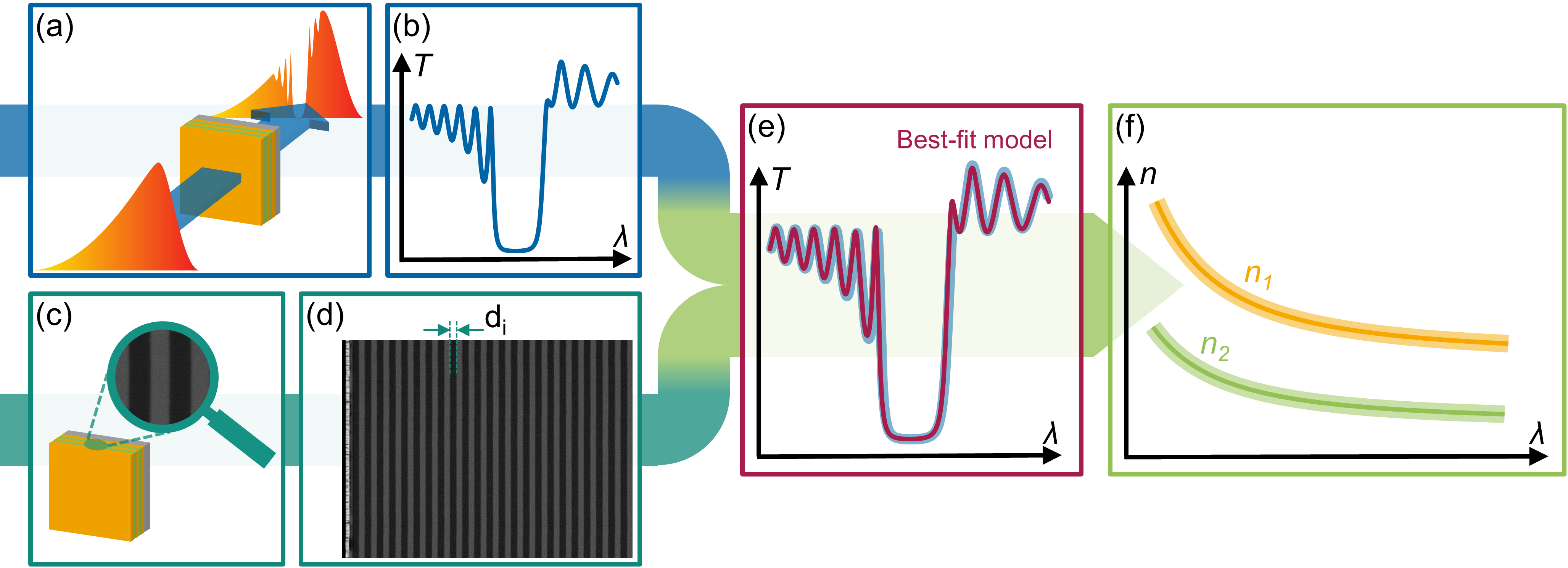}
    
    \caption{
    Schematic outline of our approach, which relies on two measurements of a thin-film heterostructure: (a) First, a spectrometric transmittance (or reflectance, not depicted) measurement is carried out. (b) From this, we acquire the optical response of the sample. (c) Second, a cross-sectional SEM (or TEM/AFM, not depicted) micrograph. (d) From this, we extract the individual layer thicknesses $d_i$. (e) These two measurements are used for a TMM-based best-fit modeling approach. (f) Finally, from this fit, we infer both refractive indices, $n_1$ and $n_2$, as well as their respective uncertainties (using a Monte-Carlo-type propagation of the measurement uncertainties from (a)--(d)).}
    \label{fig:Intro}
\end{figure*}

In this study, we introduce updated refractive index values for high-purity MBE-grown GaAs and \AlGaAsx{} in the spectral range from \SIrange{2}{7.1}{\um}, measured simultaneously via a novel, versatile method (see Fig.~\ref{fig:Intro}). While the method is adaptable to different measurement devices, the presented results leverage two widely-available instruments, a Fourier-transform infrared (FTIR) spectrometer and a field-emission scanning electron microscope (SEM), to probe a state-of-the-art GaAs/\AlGaAsx{} DBR. This DBR, with a center wavelength of approx. \SI{4.45}{\um}, has ultra-low absorption and background doping~\cite{Winkler2021Mid-infraredPpm, Truong2022Transmission-dominated000}, making it an ideal specimen for probing material optical properties. A curve-fitting routine, based on the transmission matrix method (TMM), is used to infer both refractive indices, where each material's dispersion is captured by a semi-empirical model~\cite{Afromowitz1974}. Measurement uncertainties are propagated to the final $n$ results using a Monte-Carlo-type method.

\section{Model and Theory}
\subsection{Transmission Matrix Method}
\label{sec:TMM}
A well-established means of modeling the optical response (reflectance $R$ and/or transmittance $T$) of an optical multilayer structure is the TMM~\cite{Born1999PrinciplesOptics,Byrnes2021MultilayerCalculations,Luce2022TMM-FastTutorial}. In this approach, each a $2 \times 2$-matrix represents each interface and the propagation through a given material of refractive index $n$~\cite{Byrnes2021MultilayerCalculations}, effectively relating the reflected/transmitted wave to the incoming wave via the complex reflection/transmission amplitude coefficients.
The TMM calculates the optical response of a multilayer based on the individual physical layer thicknesses $d_i$ and their respective refractive indices $n_i$.

In principle, the TMM can incorporate absorbance within the layers via the complex refractive index $\eta = n+i\kappa$, with the extinction coefficient $\kappa = \lambda_0\alpha / 4\pi$ and $\alpha$ the absorption coefficient. However, previous studies have shown that the absorbance of our structure is \SI{<10}{\ppm} (i.e., $<\SI{10e-6}{}$)~\cite{Cole2016, Winkler2021Mid-infraredPpm, Truong2022Transmission-dominated000}, which is below both the observed uncertainty in FTIR imaging (see Fig.~\ref{fig:T_spectrum_fit_and_FTIR_errors}b) as well as our targeted propagated relative uncertainty of c.~\SI{e-4}{}. Hence, we model both materials as transparent (i.e., with $\kappa=0$).

Due to its reliance on matrix multiplications, numerical calculations based on the TMM can be implemented very efficiently. We use the recently-developed \texttt{tmm-fast} package~\cite{Luce2022TMM-FastTutorial}, implemented in Python~3. This highly-optimized implementation allows for low computation times, even for many repeated calculations, as is necessary for the non-linear least-squares curve fitting routines we use. We adapt the \texttt{tmm-fast} algorithm to accommodate the treatment of thick layers where interference can be neglected, such as in the case of the MBE seed wafer, following the method described in Ref.~\cite{Byrnes2021MultilayerCalculations}.

\subsection{Refractive Index Model}
As our method involves a non-linear least-squares fit over a broad wavelength range (\SIrange{2}{7.1}{\um}), dispersion, i.e., the change of the refractive index with respect to wavelength, $\mathrm{d}n/\mathrm{d}\lambda$, must be taken into account. 
In principle, several different approaches to modeling refractive indices exist, roughly divided into empirical, semi-empirical, and theoretical models.
The choice of an appropriate model depends on several considerations. These include the electronic and optical properties of the material in the wavelength range of interest.

For GaAs/\AlGaAsx{}, several other models have been used to obtain $n(\lambda)$. Empirical models are predominantly derived from the well-known Sellmeier equation~\cite{Sellmeier1871ZurSubstanzen}. The latter was used, e.g., by Skauli et al., together with another model by Pikhtin and Yas'kov, to obtain the refractive index of GaAs in a range \SIrange{0.97}{17}{\um}~\cite{Skauli2003ImprovedOptics}.
Owing to a high number of free parameters (e.g., 7 parameters per material in the case of both models in~\cite{Skauli2003ImprovedOptics}), these models can capture variations in measurement data well. While this is generally desired, there is a potential for overfitting, as many free parameters allow for the model fit to extract residual random and/or systematic variations in the data points used for regression analysis.

In this study, we use a semi-empirical model developed specifically for GaAs and \AlGaAsx{}~\cite{Afromowitz1974}, given by the expression
\begin{equation}
\label{eq:afromodel}
  \begin{split}
    n^2(E)-1 &=\frac{\eta}{\pi} \left[ \frac{E^4_f-E^4_\Gamma}{2}+\left( E^2_f-E^2_\Gamma \right) E^2+ \right. \\
    &\left. \ln \left( \frac{E^2_f-E^2}{E^2_\Gamma-E^2} \right)E^4 \right]
  \end{split}
\end{equation}
where $E=hc/\lambda$ is the photon energy, $E_\Gamma$ is the bandgap energy. $E_f$ and $\eta$ are to be determined empirically, effectively approximating the interband optical transitions as a function $\epsilon_2=\eta E^4$ for $E_\Gamma<E<E_f$ ($\epsilon_2=0$ otherwise).

The lowest direct bandgap energy for pure GaAs and AlAs are given by $E_{\Gamma,\mathrm{GaAs}}=\SI{1.424(1)}{\eV}$ (measured at a temperature of \SI{297}{\kelvin})~\cite{Sell1974ConcentrationEV} and $E_{\Gamma,\mathrm{AlAs}}=\SI{3.018}{\eV}$ (measured at room temperature, given without uncertainty)~\cite{Casey1978HeterostructureLasers}, respectively. Aspnes et al. showed that for arbitrary $x$, $E_\Gamma$ is well-approximated by
\begin{equation}
\label{eq:EGamma-arb_x}
  E_\Gamma(x) = 1.424 + 1.721 x - 1.437 x^2 + 1.310 x^3
\end{equation}
which allows us to determine $E_\Gamma(x)$ based on an independent measurement of $x$.
As shown in~\cite{Afromowitz1974}, the expression in (\ref{eq:afromodel}) closely reproduced the refractive index of GaAs for wavelengths from \SIrange{0.895}{1.7}{\um}, with similar results for AlAs. We note that Ref.~\cite{Afromowitz1974} also gives an interpolation scheme that can be used to approximate $n$ for different $x$ based on our results below.

In an effort to avoid bias introduced by the model selection, we compare these results to a fit using the so-called Single Effective Oscillator (SEO) model~\cite{Wemple1971BehaviorMaterials}, which, while not suited for the NIR, approximates $n$ exceptionally well for $E \ll E_{\Gamma,\mathrm{GaAs}}$ ($\lambda\geq \SI{2}{\um}$)~\cite{Palmer2002Mid-infraredAlAs, Afromowitz1974}. According to this model~\cite{Wemple1971BehaviorMaterials}, the refractive index $n$ of a crystalline material at a photon energy $E$ far below the bandgap is well approximated by
\begin{equation}
\label{eq:SEO}
  n^2(E)-1 = \frac{E_0 E_d}{E_0^2-E^2}
\end{equation}
In this approximation, the interband optical transitions are considered as a single dipole transition at energy $E_0$, with an effective oscillator strength of $\pi E_d/2$. Hence, the dispersion of GaAs/\AlGaAsx{} in the MIR range can be adequately modeled fitting only two free parameters, $E_0$ and $E_d$, as was shown in Ref.~\cite{Palmer2002Mid-infraredAlAs}.

\subsection{Monte Carlo Error Propagation}
We use a Monte-Carlo-type routine to propagate the measured uncertainties to the best-fit model parameters. This approach is robust to common systematic measurement errors because the refractive indices are tightly constrained by the measured quantities.

For the purpose of uncertainty propagation, the model can be described as a function
\begin{equation}
\label{eq:MCprop}
 T = \mathrm{TMM}(d_i, n_{\mathrm{Subst}}, n_{\mathrm{GaAs}}, n_{\mathrm{AlGaAs}})
\end{equation}
denoting the transmittance $T$ as a function of the layer thicknesses $d_i$ and refractive indices $n$. The fitted parameters $n_{\mathrm{GaAs}}$ and $n_{\mathrm{AlGaAs}}$ are described according to Eqs.~(\ref{eq:afromodel}--\ref{eq:SEO}). Here, $T$, $d_i$, and $n_{\mathrm{Subst}}$ represent measured quantities with associated uncertainties. Hence, our method involves using measured quantities not only in the data the model is fit to (the $T$ measured via FTIR), but also in the parameters used to seed the model ($d_i$ and $n_\mathrm{Subst}$). Since the TMM represents a complicated model function, and a non-linear least-squares fit routine is used, error propagation to the best-fit parameters is not straightforward and no standardized procedure exists. 

To overcome this challenge, we use a Monte-Carlo approach to calculate the propagated uncertainty: For each run, we randomly pick a certain realization of the measurement values (layer thicknesses $d_i$ and transmittance values $T(\lambda)$) from distributions based on their respective measurement uncertainties (as given in Fig.~\ref{fig:SEM_layers_and_errors}b and \ref{fig:T_spectrum_fit_and_FTIR_errors}b).
We repeat this process many times, yielding slightly different sets of parameters that vary within the measurement uncertainties.
We then use each set for the same curve-fitting exercise, in which the dispersion parameters for both materials are recorded. This results in datasets for each fit parameter that also show variation as a consequence of the uncertainty in the measurement data.

Finally, for each set of parameters, the refractive index $n(\lambda)$ is calculated for both materials. It is of note that the uncertainty in $n$ is much smaller than what would result from Gaussian error propagation because the fitted parameters show appreciable covariance with the input parameters taken from measurement and literature values. This is best seen when comparing the uncertainties resulting from fitting Eq.~(\ref{eq:afromodel}) vs. Eq.~(\ref{eq:SEO}): The former requires literature and measurement values for $E_{\Gamma}$ and $x$ with associated additional uncertainties when compared to the latter. Still, this leads to a negligible difference in the observed standard uncertainty of $n$.

\section{Experiment}
The basic idea of our approach is the following:
On the one hand, the optical response, in our case the transmittance $T$, of a transparent optical system can be accurately modeled via TMM if the physical structure (i.e., layer thicknesses) and the associated refractive indices are known. On the other hand, the transmittance is accessible to optical probing. As a result, knowing both the measured optical response as well as the material composition and layer thicknesses of an optical structure allows us to determine the refractive indices. This means that the refractive indices can be uniquely defined for a two-material structure that exhibits a characteristic optical response in the wavelength range of interest.

While this work details results for GaAs/\AlGaAsx{}, the underlying measurement principle (combining knowledge of the optical response with measurements of the physical structure of a thin-film multilayer to simultaneously obtain the refractive indices of both materials) is much broader in its applicability, as is shown in Fig.~\ref{fig:Intro}. It can be adapted for different samples, measurement devices, and spectral ranges. In this study, data collection consists of two independent measurements:
\begin{enumerate*}[label=(\roman*)]
  \item obtaining a photometrically accurate transmittance $T$ spectrum via Fourier-transform infrared (FTIR) spectrometry
  \item measuring individual physical layer thicknesses $d_i$ via calibrated SEM metrology.
\end{enumerate*}
As explained in~\ref{sec:TMM}, the TMM relates the optical layer thicknesses to the optical response of a layered system.

\subsection{Fabrication and Description of the GaAs/\AlGaAsx{} Sample}
We designed the sample under test to serve as half of a distributed Bragg reflector (DBR) structure centered at $\lambda_\mathrm{d}=\SI{4.5}{\um}$. 
Nominally, this structure consists of 22.5 periods of \AlGaAsx{}/GaAs layer pairs, with an optical layer thickness of $\lambda_\mathrm{d}/4$, terminated by a single $\lambda_\mathrm{d}/8$-thickness GaAs cap (46 layers total) to avoid exposure of \AlGaAsx{} to air, preventing oxidation. The target AlAs mole fraction in the \AlGaAsx{} layers was $x=0.92$.

Based on the above design, a DBR specimen is deposited via MBE on a [001]-oriented semi-insulating (GaAs with As anti-site defects) GaAs seed wafer with a diameter of \SI{15}{\cm} and a nominal thickness of \SI{675(25)}{\um}. From this as-grown structure (seed wafer plus MBE-deposited heterostructure), we cleave a rectangular die of ~\SI[parse-numbers=false]{2\times 2}{\cm\squared}, which serves as the sample for all parts of this study. After growth, the AlAs mole fraction $x$ is estimated to be \SI{92.9(30)}{\percent} based on x-ray diffraction (XRD) measurement performed by the epitaxial material supplier. Here, the uncertainty represents a conservative estimate based on data from~\cite{Bertness2000AlGaAsReflectance,Bertness2006Standard260-163}. In our analysis, we assume that the mole fraction $x$ is the same for all \AlGaAsx{} layers and that interfaces between the layers are abrupt (i.e., that $n$ changes over a distance $\ll \lambda$).

Recently, a two-mirror cavity using 44.5-period DBR mirrors fabricated with material from the same growth run demonstrated a Finesse of \SI{230000}{}, corresponding to a per-mirror excess loss (absorption plus scatter) of $\SI{4.27(52)}{\ppm}$. From this an average extinction coefficient $k<10^{-6}$ was extracted, suggesting background doping levels of $\leq \SI{e-14}{\per\cm\cubed}$~\cite{Truong2022Transmission-dominated000}. This high purity makes the multilayer structure an ideal specimen to extract the refractive index $n$ of both materials from an optical measurement.

\subsection{Transmission Spectrum via Fourier-Transform Infrared Spectrometry}
We obtain transmittance $T$ spectra using a commercially available vacuum FTIR spectrometer (Bruker Vertex 80v). As described below, the necessary photometric accuracy was obtained by thorough calibration and rigorous control of measurement conditions.

\subsubsection{Measurement Conditions and Parameters}
To achieve excellent photometric accuracy, we thoroughly characterize the FTIR device and optimize the measurement parameters.
From these efforts, we find that a SiC globar-type light source with a \SI{2}{\mm} aperture is optimal. The aperture size is chosen to maximize throughput while obtaining a well-collimated beam (excluding influence on resolution due to a large source diameter).
We use a standard KBr beamsplitter for optimal modulation efficiency over the wavelength region of interest.
The scan speed of the moving interferometer mirror is set to \SI{10}{\kHz} (corresponding to a physical scan speed of c.~\SI{0.32}{\cm\per\s}), limited by the RF-frequency bandwidth of the employed DLaTGS pyroelectric detector.
We select this detector to minimize systematic errors due to nonlinear detector response, owing to its excellent linearity and flat detectivity $D^*(\lambda)$ over our measurement range~\cite{Theocharous2006DetectorsUse,Theocharous2008AbsoluteDetector}.
We verified the wavelength calibration of the FTIR by comparing the spectrum of a polystyrene filter with a manufacturer-supplied calibration curve.
We perform all measurements after evacuation (\SI{2.21}{\milli\bar}). We find that optimal stability and repeatability of the above configuration is achieved \SI{>10}{\hour} after starting evacuation and switching on the light source. Possible causes for this are thermalization (light source, detector) or a gradual improvement in evacuation. It is of note that this is well above the stabilization time recommended by the manufacturer (\SI{4}{\hour}), which we attribute to the aging of our device.
Because of the detector's moderate $D^*$ and the resulting single shot signal-to-noise ratio (SNR), each individual measurement (duration of \SI{15}{\min} each) is the average of 256 individual interferograms.

\begin{figure}[htbp]
\includegraphics{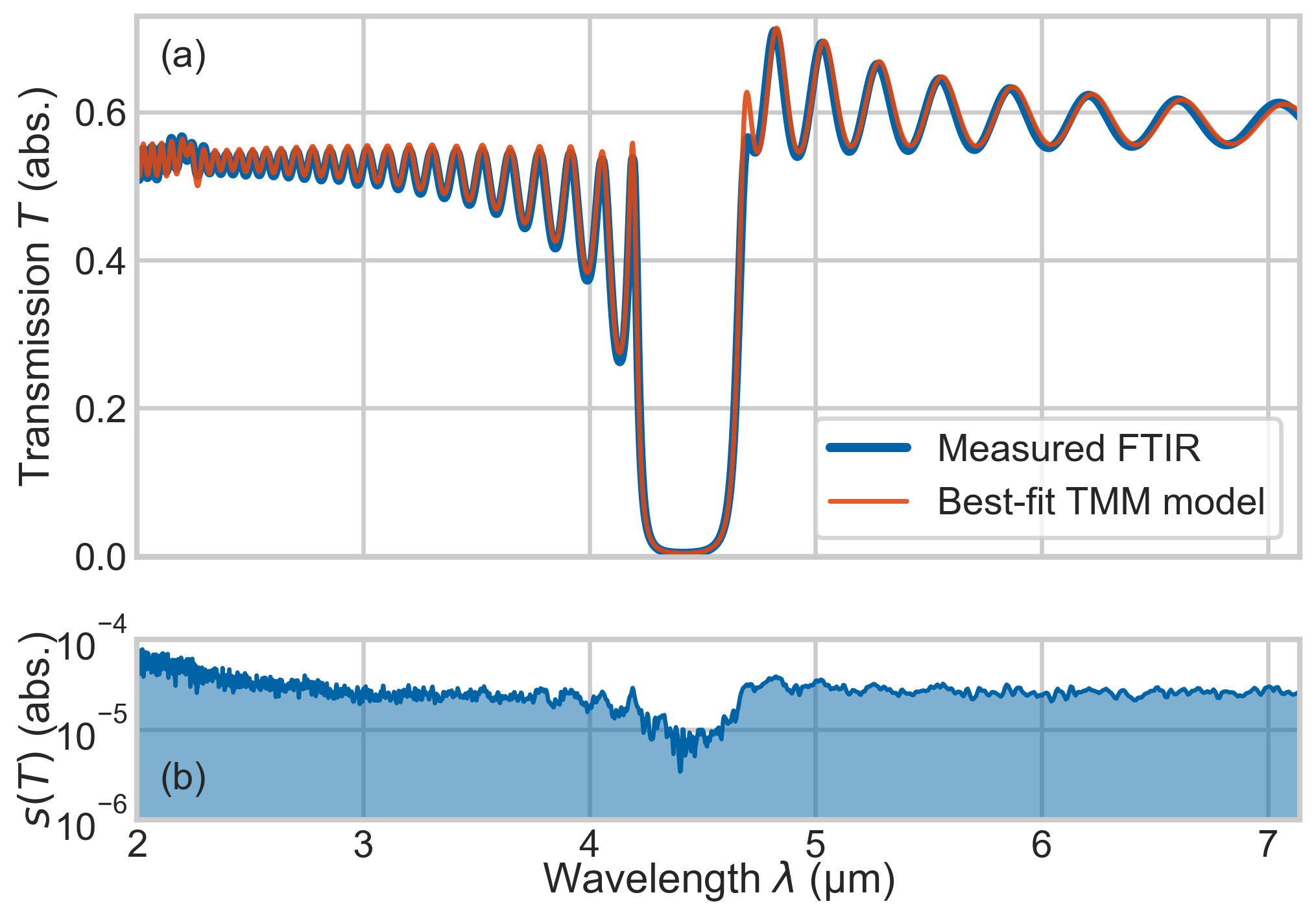}
\caption{\label{fig:T_spectrum_fit_and_FTIR_errors} (a) Transmittance spectra of the measured multilayer structure. As-measured in FTIR (blue) and the best-fitted model (orange) obtained when calculating a nonlinear least-squares regression with the TMM model based on the physical layer thicknesses extracted from SEM. (b) Statistical $1s$ standard uncertainty of the FTIR measurements.}
\end{figure}

We record all measurements as double-sided interferograms. That way, the phase spectrum is available at the same resolution as the power spectrum for the FT process, avoiding photometric errors from phase correction of single-sided interferograms~\cite{Chase1982PhaseFT-IR}.
Subsequently, we Fourier-transform each measurement to a spectrum with \SI{5}{\per\cm} resolution, using a Norton-Beer medium apodization and Mertz phase correction in the process, previously shown to yield optimal photometric accuracy~\cite{Griffiths2007FourierSpectrometry}.
To obtain the transmittance spectrum of the DBR, we ratio the sample spectrum against a background spectrum which is measured and evaluated under identical conditions immediately before.
We repeat the background (sample size $i=14$) and sample ($i=8$) measurements several times to estimate the associated type A uncertainties for each series over our entire measurement range \SIrange{1400}{5000}{\per\cm} (see Fig.~\ref{fig:T_spectrum_fit_and_FTIR_errors}b).

We stabilize the sample temperature at $T_s=\SI{22(1)}{\celsius}$ using a custom-built optomechanical mount with a single-stage TEC element to exclude fluctuations due to thermal drift. We carefully aligned the sample to normal incidence (overlapping the reflected light with the incident beam at the entrance of the sample chamber), with an estimated error of \SI{<0.3}{\degree}.

\subsubsection{Temperature-resolved measurements}
\begin{figure}[htbp]
\includegraphics[width=8.6 cm]{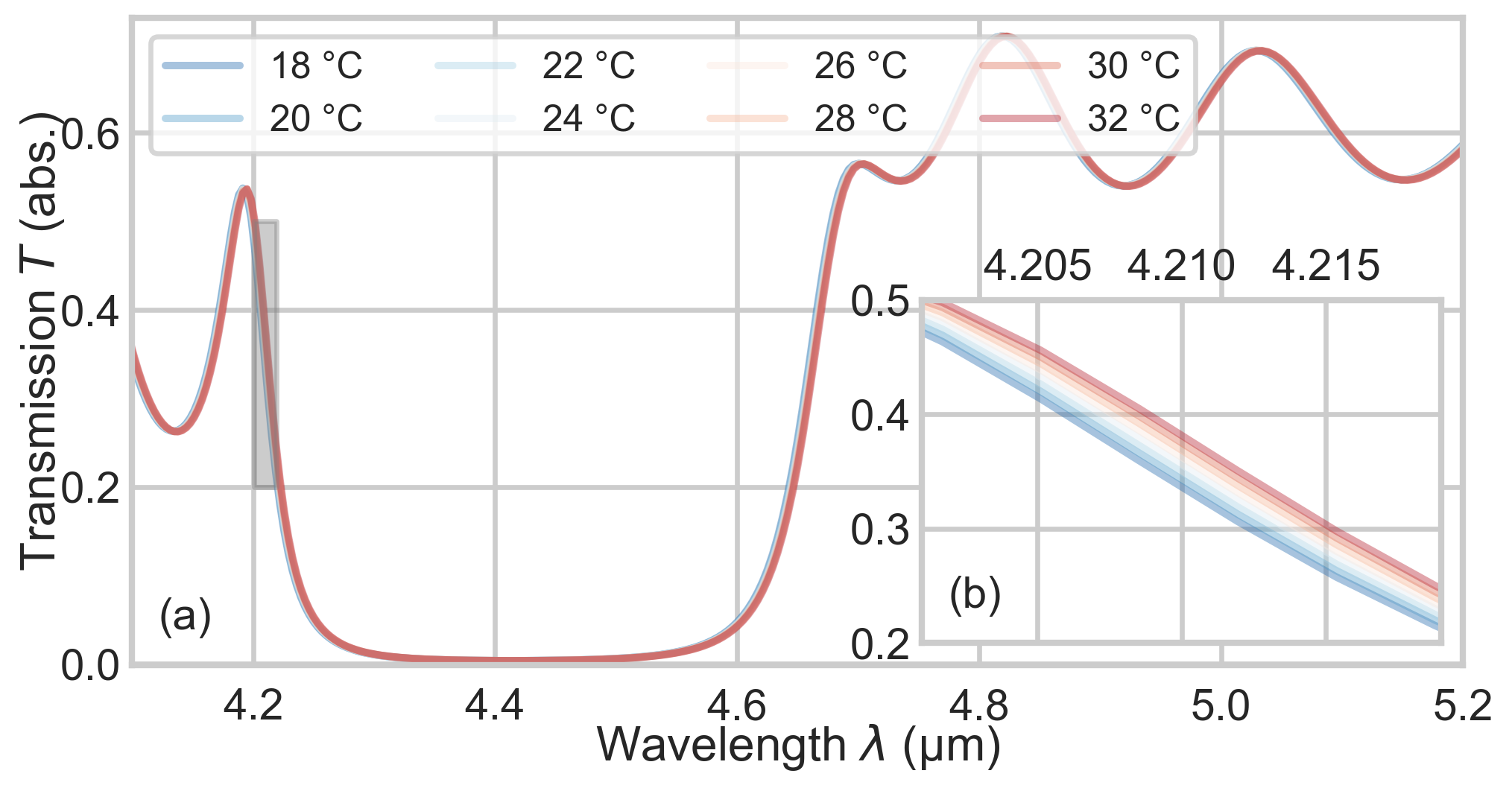}
\caption{(a) Temperature resolved measurements spanning \SIrange{18}{32}{\celsius}, including the measurement used for retrieving the refractive indices at \SI{22}{\celsius}. (b) As systematic shifts are barely visible in the main figure, we show a zoom of the shaded area in (a). The shift of approx. \SI{3}{\nm} at \SI{4210}{\nm}} due to changes in temperature is clearly visible.
\label{fig:T_spectrum_temp}
\end{figure}

The aforementioned TEC-stabilized mount allowed us to cool/heat the sample in a limited range (\SIrange{18}{32}{\celsius}).
The temperature stability is better than \SI{\pm 0.1}{\kelvin}, while the accuracy is limited by the thermistor (including calibration error) to \SI{\pm 1}{\kelvin}. Temperature readings are given in Fig.~\ref{fig:T_spectrum_temp}. As discussed in~\cite{Skauli2003ImprovedOptics}, such temperature dependent measurements can be used to extract $\mathrm{d}n/\mathrm{d}T_s$ and $\mathrm{d}^2 n/\mathrm{d}T_s^2$, as the linear thermal expansion coefficient of GaAs/\AlGaAsx{} is well known~\cite{Skauli2003ImprovedOptics}. Accurate extraction of these parameters would require a wider range of temperatures, beyond the capabilities of our custom-built mount.

\subsection{Scanning Electron Microscope Measurements}
We measure the individual thickness of each GaAs/\AlGaAsx{} layer using a scanning electron microscope (Zeiss Supra 55 VP) providing a cross-section of the same sample used in FTIR measurements.

The described routine takes direct advantage of the excellent thickness uniformity of epitaxial DBRs, which was demonstrated to be less than \SI{0.41(5)}{\nm} RMS over a \SI{4}{\cm} diameter sample in~\cite{Koch2019ThicknessInterferometry}. This allows us to extract individual measurements from several thousand high-resolution line scans to calculate sample statistics, drastically reducing the standard uncertainty.

\subsubsection{Sample Preparation}
Prior to preparing the cross-section, we deposit a gold layer on top of the heterostructure using a table-top sputtering device (Leica EM SCD050 with EM QSG100) to avoid contrast issues at the air-GaAs boundary in SEM metrology (see Fig.~\ref{fig:SEM_img_and_line}).
To reveal the cross-section of the multilayer, we cleave the die within the c.~ \SI{2}{\mm} spot probed in FTIR spectrometry using a standard diamond scribe.
As we use a monocrystalline superlattice structure this results in a cross-section perpendicular to the front surface (flatness was confirmed using a secondary electron detector in SEM). We mount the sample on a \SI{90}{\degree} sample holder to exclude systematic errors (tilt angle and vibration).

As \AlGaAsx{} is known to form oxide layers, we take care that the sample is exposed to the atmosphere prior to SEM imaging for no more than \SI{15}{\minute}. According to prior studies~\cite{Reinhardt1996OxidationAir}, this will cause an oxide layer of c.~\SI{1}{\nm} thickness for our sample. We find this to be negligible compared to the mean interaction depth of backscattered electrons, simulated to be $\gg\SI{1}{\nm}$ at \SI{10}{\kilo \eV} (using CASINO v2.51~\cite{Drouin2007CASINOUsers}). Hence, this thin oxide film does not affect SEM imaging using a backscattered electron detector.

\subsubsection{Cross-Sectional Measurement}
\begin{figure}[htbp]
  \includegraphics{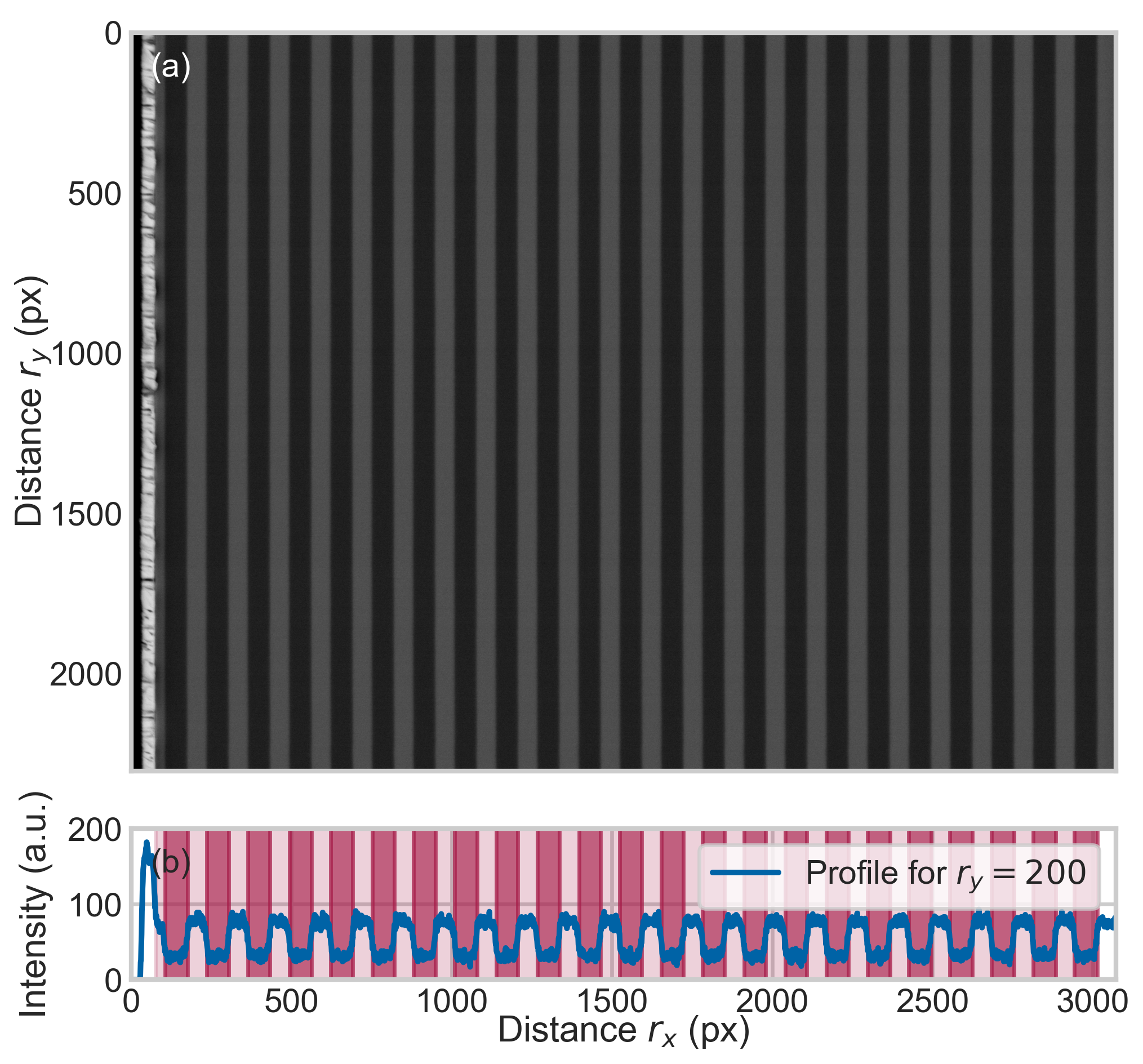}
  
  \caption{\label{fig:SEM_img_and_line}
  (a) One of a total of four SEM images of the DBR under test. The leftmost layer is gold, coated after FTIR to exclude boundary effects in SEM metrology. The size of this micrograph is c.~\SI{17.3 x 12.9}{\um}, resulting from a calibrated pitch of \SI{5.6278(17)}{\nm\per\pixel} 
  (b) The typical profile of a single row (blue) with extracted layer thicknesses (red).}
\end{figure}

We load the mounted sample on the SEM's multi-sample holder together with a certified calibration standard (EM-TEC MCS-0.1CF).
After evacuation of the system (c.~\SI{5e-5}{\milli\bar}), we perform all measurements at a beam energy of \SI{10}{\kilo\electronvolt}. All images used in this evaluation were obtained at a working distance of $\approx \SI{9.5}{\mm}$.
Prior to calibration and measurement, we set a nominal magnification of c.~6600, so that the complete superlattice structure is captured in a single image.
We use the standard's \SI{1}{\um} grid (17 lines with \SI{1}{\um} pitch, certified total length of \SI{16.0100(48)}{\um}) to calibrate the SEM at the chosen magnification.
We image the standard with the above settings after optimizing the stigmator and focus to obtain an undistorted micrograph (see Fig.~\ref{fig:SEM_img_and_line}a).
Following the calibration step, we use the translation stage of the SEM to position the cross-section of our heterostructure below the electron gun. In this step, we also corrected for a slight height difference between the standard and sample by moving the cross-section into focus (at identical working distance etc.). Thereby, the heterostructure is imaged under identical conditions as the standard at 4 distinct positions.

\subsubsection{Evaluation}
For both the standard and the sample, we use a custom Python~3 script to extract the interface positions according to the following steps:
\begin{enumerate*}
  \item loading the picture as a 2D array, where every entry is assigned the 8-bit grayscale value of the corresponding pixel,
  \item slicing the image row-by-row to get a line profile (see Fig.~\ref{fig:SEM_img_and_line}b),
  \item estimating the interface positions by finding the extrema of the first derivative of the row-wise profile,
  \item dividing each single row profile into smaller intervals around the estimated interface positions so that each interval contains a single interface,
  \item for each interface, fitting an error function and using the inflection point of the fitted curve as our estimate for the interface position (using the error function is justified by the fact that it is the convolution of an electron beam, approximated by a Gaussian profile, and the abrupt material interface, approximated by a step function),
  \item layer thicknesses are obtained by subtracting subsequent estimates for the interface positions from each other
  \item repeating the above procedure for all rows in the image, we find the mean and $1s$ standard uncertainty of the mean for each layer $s(d_i)$.
\end{enumerate*}

In the case of the sample, we repeat these steps for all four images. In this procedure, we exclude significant systematic errors due to sample tilt by testing for systematic changes in extracted interface positions.

From the evaluation of the reference standard, we establish a calibrated distance per imaged pixel. For that, we measure the certified distance of the standard to be \SI{2844.807(10)}{\pixel}. This results in a calibrated pitch of \SI{5.6278(17)}{\nm\per\pixel}. The propagated uncertainty of \SI{0.03}{\percent} is dominated by the certified uncertainty of the standard, as the relative statistical uncertainty is \SI{<4}{\ppm}.

\begin{figure}[htbp]
  \includegraphics{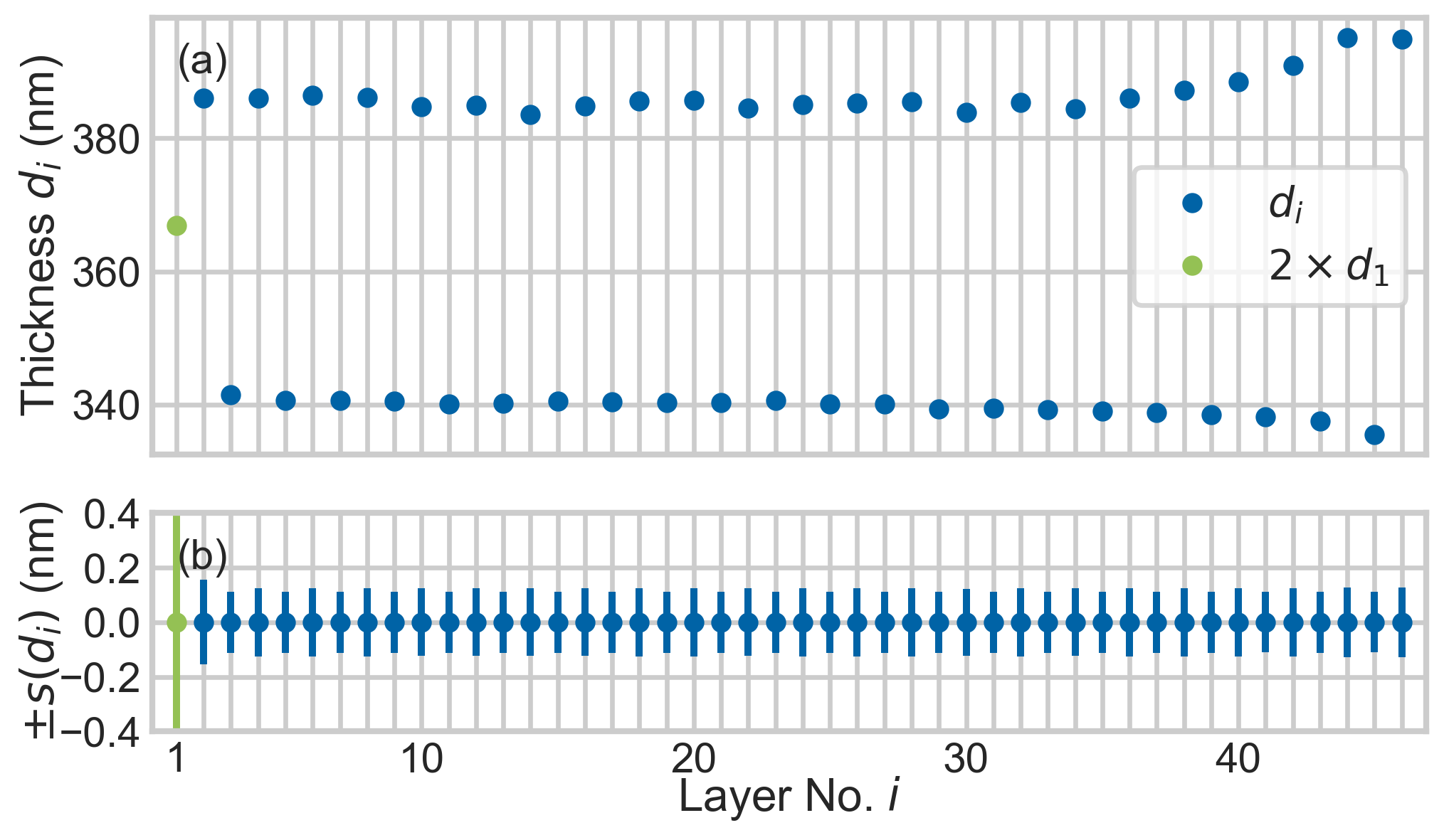}
  
  \caption{\label{fig:SEM_layers_and_errors}
  (a) Measured layer thicknesses, resulting from averaging row-wise measurements. Note that the thickness of the 1/8-wave cap, $d_1$, (green) was multiplied by a factor of 2. (b) Error bars showing $1s$ standard uncertainty for the mean values given in (a).
  }
\end{figure}

From four SEM images of the GaAs/\AlGaAsx{} heterostructure, we extract the mean and standard uncertainty of the individual layer thicknesses in units of pixels.
We multiply these values by the calibrated pitch to assign physical layer thicknesses in units of nanometers, as summarized in Fig.~\ref{fig:SEM_layers_and_errors}.

The propagated relative uncertainty ($<\SI{4e-4}{}$ for all $\lambda_\mathrm{d}/4$ layers) is dominated by the uncertainty of the calibration standard, except for the topmost $\lambda_\mathrm{d}/8$ GaAs layer, which exhibits a relative  uncertainty of \SI{2.2e-3}{}. This is caused by the Au/GaAs interface, where the data quality suffers from remaining structures in the Au cross section that could not be eliminated.

\section{Results}
\label{sec-results}
To obtain the refractive indices of GaAs/\AlGaAsx{} from the above measurements, we use a non-linear least-squares fitting approach. In that process, we seed a TMM model based on the layer thicknesses obtained via SEM and model the refractive indices according to the models in Eq.~(\ref{eq:afromodel}) and~(\ref{eq:SEO}). We then fit the resulting curve to the photometrically-accurate FTIR transmittance $T$ spectrum (see Fig.~\ref{fig:T_spectrum_fit_and_FTIR_errors}a) to obtain values for the free parameters, effectively resulting in values for the refractive indices of GaAs/\AlGaAsx{}.

\begin{figure}[htbp]
  \includegraphics{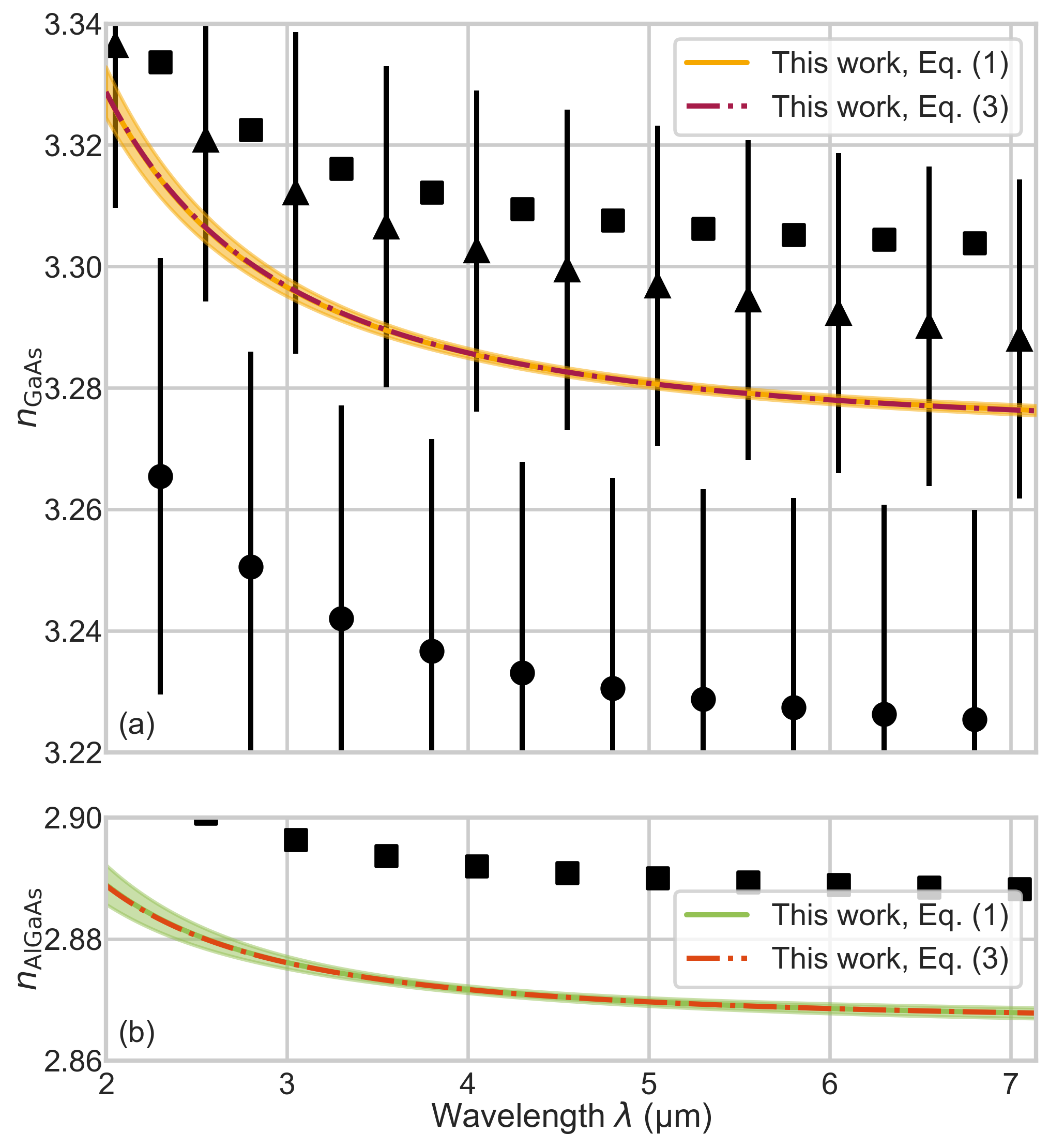}
  
  \caption{\label{fig:final_n_results}
  Final results for the refractive indices, compared to values from literature. Error bands/bars for our results obtained from fitting Eq.~(\ref{eq:afromodel}) and for values from Ref.~\cite{Skauli2003ImprovedOptics} (triangles, Pikhtin model) are given as fourfold standard uncertainty $4s(n)$, while the error bars for Ref.~\cite{Palmer2002Mid-infraredAlAs} (circles) are given as $1s(n)$. Values for Ref.~\cite{Afromowitz1974} (squares) represent an extrapolation from the NIR with unknown uncertainty. Error bands for our fit of Eq.~(\ref{eq:SEO}) are omitted because they are almost identical to those for Eq.~(\ref{eq:afromodel}). (a) Refractive index of GaAs. (b) Refractive index of \AlGaAs{}.
  }
\end{figure}

To infer accurate mean values and standard uncertainties for both refractive indices, we use the described Monte-Carlo-type uncertainty propagation. 
This involves running the fit routine 1000 times for each model. Before each fit, we randomly pick the transmittance values $T(\lambda)$, the individual layer thicknesses $d_i$, and, in the case of Eq.~\ref{eq:afromodel}, the AlAs content $x$ and literature values for $E_{\Gamma}$, from normal distributions representing their respective propagated uncertainty. The resulting distributions, which give the mean and standard uncertainty for $n$, also closely follow a normal distribution.
Subsequently, we use each set of Monte-Carlo parameters to calculate the refractive index, where the variation of these calculations results in mean values and standard uncertainties for both indices $n(\lambda)$, which we show in Fig.~\ref{fig:final_n_results}. We report model parameters reproducing our results in Tab.~\ref{tab:fit-results} for ease of computation. The largest relative standard uncertainty is observed at the lower-wavelength end of our spectrum. Using Eq.~(\ref{eq:afromodel}) with Eq.~(\ref{eq:EGamma-arb_x}), these values are 
$s(n_{\mathrm{GaAs}})/n_{\mathrm{GaAs}} \leq \SI{3.2e-4}{}$ and $s(n_{\mathrm{AlGaAs}})/n_{\mathrm{AlGaAs}} \leq \SI{2.8e-4}{}$, whereas uncertainties for Eq.~(\ref{eq:SEO}) are $\leq \SI{3.3e-4}{}$ and $\leq \SI{2.9e-4}{}$ for GaAs and \AlGaAs{}, respectively. We report these as the uncertainty for $n$ when calculated with parameters in Tab.~\ref{tab:fit-results} over the whole wavelength range.

\begin{table}[b]
\caption{\label{tab:fit-results}%
Parameters to calculate the refractive indices for both materials according to our results for models given by Eq.~(\ref{eq:afromodel}) and Eq.~(\ref{eq:SEO}).
}
\begin{ruledtabular}
\begin{tabular}{lllll}

Material    & Model                     & Variable & Value \\
\colrule
GaAs        & Eq.~(\ref{eq:EGamma-arb_x})   & $x$       & 0 \\
            & Eq.~(\ref{eq:afromodel})      & $\eta$    & 0.170481 \\
            &                               & $E_f$     & 4.36144 \\                   
            & Eq.~(\ref{eq:SEO})            & $E_0$     & 3.21318 \\
            &                               & $E_d$     & 31.1845 \\
\colrule
\AlGaAs{}   & Eq.~(\ref{eq:EGamma-arb_x})   & $x$       & 0.929 \\
            & Eq.~(\ref{eq:afromodel})      & $\eta$    & 0.0393107 \\
            &                               & $E_f$     & 5.90697 \\        
            & Eq.~(\ref{eq:SEO})            & $E_0$     & 4.64100 \\
            &                               & $E_d$     & 33.4811
\end{tabular}
\end{ruledtabular}
\end{table}

\section{Conclusion and Outlook}
In summary, we report the simultaneous measurement of refractive index $n$ values for both pure GaAs and \AlGaAs{} in the spectral range from \SIrange{2}{7.1}{\um} at room temperature ($T_s=\SI{22+-1}{\celsius}$), probing a superlattice DBR, grown via MBE with exceptionally low background doping~\cite{Truong2022Transmission-dominated000}.

We obtain these measurements via a general, highly-adaptable method, that allows us to measure the refractive indices of two materials simultaneously by probing a thin-film heterostructure. The approach can be used with crystalline and amorphous dielectric multilayers. These multilayers bear the advantage that one of the materials is not exposed to the atmosphere, easing the characterization of oxidizing materials such as \AlGaAsx{}, avoiding complicated schemes~\cite{Reinhardt1996OxidationAir,Papatryfonos2021RefractiveRegion}. This enables routine refractive index measurements of current and future optical materials in their transparent range.
As outlined in Fig.~\ref{fig:Intro}, the individual steps involve acquiring the optical response and accurately measuring the layer thicknesses. The exact layer thicknesses are measured in our process. In our method, we accurately measure the thickness of each layer. Consequently, the sample does not have to be a DBR, provided that the transmittance spectrum has broadband characteristics, which remain undistorted by the resolution of the spectrometer in use. Many multilayer structures, such as antireflection (AR) or broadband high-reflectivity (BBHR) coatings, typically meet this criterion. A subsequent non-linear least-squares fit results in accurate refractive index values over a broad wavelength range, capturing material dispersion with suitable empirical, semi-empirical, or theoretical models for both refractive indices. Compared to other approaches, the presented routine realizes high levels of accuracy and precision, while reducing experimental complexity and relying on widely-available devices. We do not require specialized and cost-intensive optical setups, such as spectroscopic ellipsometers~\cite{Fujiwara2007SpectroscopicEllipsometry}.
The evaluation step avoids intricate extrapolation routines, which are needed in the fringe pattern analysis used for FTIR refractometry~\cite{Kaplan1998FourierRefractometry, Skauli2003ImprovedOptics}.

In the present study, we acquire a photometrically-accurate transmittance spectrum via FTIR spectrometry and an accurate measurement of individual layer thicknesses via calibrated SEM metrology. For both measurements, control of systematic errors was necessary but achieved by simple means such as careful alignment and temperature stabilization.
Subsequently, we perform a non-linear least-squares fitting routine based on TMM, where the dispersion $\mathrm{d}n/\mathrm{d}\lambda$ was captured according to two different models for each refractive index. Finally, we use the best-fit results to obtain the refractive indices for GaAs/\AlGaAs{} over the entire wavelength range.
Propagation of measurement uncertainties via a Monte-Carlo approach suggests relative uncertainties on the order of $10^{-4}$ for both materials, achieving good agreement with previously-published results. Differences to the values published in~\cite{Afromowitz1974} are explained by the fact that this data represents an extrapolation from a fit to NIR measurements. Notably, the data by Palmer et al.~\cite{Palmer2002Mid-infraredAlAs} shows the largest discrepancy, likely caused by a systematic offset due to the incorrect assumption of perfectly uniform high- and low-index layers. The discrepancy with regards to the results by Skauli et al.~\cite{Skauli2003ImprovedOptics}, which were used to model the GaAs wafer in the present study, is likely caused by different material properties, such as free-carrier concentrations~\cite{Sell1974ConcentrationEV} when comparing MBE-grown GaAs with LEC/VGF-grown GaAs wafers.

We believe that both the proposed method and the updated values for GaAs/\AlGaAs{} should find use in the design, fabrication, and characterization of active and passive optical devices. This is of special importance for the MIR, which is of high interest for applications in spectroscopy, but also a region where the optical properties of many materials are still poorly studied.

\begin{acknowledgments}
We thank Dr. Stefan Eichler at Freiberger Compound Materials GmbH for GaAs wafer characterization data, Prof. Thomas Pichler for access to his group's FTIR device, Dr. Valentina Shumakova for her help with Fig.~\ref{fig:Intro}, and Dr. Adam J. Fleisher at NIST Gaithersburg for in-depth discussions and helpful advice.

We acknowledge support from the Faculty Center for Nano Structure Research at the University of Vienna, providing the SEM. The financial support by the Austrian Federal Ministry for Digital and Economic Affairs, the National Foundation for Research, Technology and Development and the Christian Doppler Research Association is gratefully acknowledged.
\end{acknowledgments}

\bibliography{main}

\end{document}